\documentclass[referee]{mn2e}

\usepackage{graphicx}
\usepackage{amsmath}

\title[Gamma-rays from binary system PSR J1227-4853]
{Gamma-ray emission states in the redback millisecond pulsar binary system PSR J1227-4853}
\author[W. Bednarek]
{W. Bednarek \\ 
Department of Astrophysics, The University of \L \'od\'z,
ul. Pomorska 149/153, 90-236 \L \'od\'z, Poland,\\
bednar@uni.lodz.pl}
\begin{document}

\date{Accepted . Received ; in original form }

\pagerange{\pageref{firstpage}--\pageref{lastpage}} \pubyear{2015}

\maketitle

\label{firstpage}

\begin{abstract}
Long expected transition states between the rotation powered and accretion powered non-thermal emission in the millisecond pulsar binary systems have been recently observed in the case of three objects PSR J1023+0038, PSR J1824-2452, and PSR J1227-4859. Surprisingly, the transition is related to the significant change in the $\gamma$-ray flux being a factor of a few higher with the presence of an accretion disk. The origin of this enhanced emission 
seems to be related to the penetration of the inner pulsar magnetosphere by the accretion disk. We propose that the radiation processes, characteristic for the rotation powered pulsar, can co-exist with the presence of an accretion disk in the inner pulsar magnetosphere. In our scenario additional $\gamma$-ray emission is produced by secondary leptons, originated close to the acceleration gap, which Compton up-scatter thermal radiation from the accretion disk to GeV energies. The accretion disk penetrates deep into the pulsar magnetosphere allowing the matter to fall onto the NS surface producing pulsed X-ray emission. We show that the sum of the rotation powered pulsar $\gamma$-ray emission, produced by the primary electrons in the curvature process, and the $\gamma$-ray emission, produced by secondary leptons, can explain the observed high energy radiation from the redback binary pulsar PSR J1227-4853 in the state with evidences of the accretion disk.
\end{abstract}

\section{Introduction}

The popular scenario for the formation of millisecond pulsars (MSPs) postulates transfer of an angular momentum with the matter accreting onto a neutron star (NS) surface from a companion star (Alpar et al.~1982, Radhakrishnan \& Srinivasan 1982). This hypothesis has recently got strong support with the discovery of three MSP binary systems that showed transition from the rotation powered pulsar state to the accretion disk powered state. The binary system containing PSR J1023+0038 has shown at first the evidences of an accretion disk with a relatively low power  
($<10^{35}$ erg s$^{-1}$) which moved to the rotation powered pulsar with clear radio pulsations (Archibald et al.~2009). Another binary, containing PSR J1824-2452, was at first observed as a rotation powered pulsar but moved to the low luminosity accretion state (Papitto et al.~2013). In the third binary, the low luminosity disk accretion state moved to the clear radio MSP state showing also 
$\gamma$-ray pulsations (de Martino et al.~2010, Roy et al.~2014). 
Recently, another binary system 1RXS J154439.4-112820, most probably associated with $\it Fermi$ source 3FGL J1544.6-1125, has been also proposed to belong to this small class of objects 
(Bogdanov \& Halpern~2015).
Interestingly, X-ray pulsations with the period of the pulsar have been observed in PSR J1824-2452 
(Papitto et al. 2013), PSR J1023+0038 (Archibald et al.~2014) and PSR J1227-4853 (Papitto et al.~2014a). These last results indicate that the matter can reach the NS surface also at low
accretion rates corresponding to the X-ray luminosities of the order of $\sim$10$^{34}$ erg s$^{-1}$.
These three binary systems belong to the redback class which contain companion stars with the mass $\sim$0.2 M$_\odot$ accreting matter through the 1st Lagrangian point.
Thus, the formation of the accretion disks around these pulsars is expected.
It is reported that the transition between two states in the binary systems containing PSR J1023+0038 and PSR J1227-4853 is connected with the change of the  $\gamma$-ray flux which is a factor of a few larger in the state with the presence of the accretion disk (Takata et al.~2014, Xing \& Wang~2014, Johnson et al.~2015). 

Two models for the high energy processes in transiting millisecond pulsar binary systems have been recently considered to explain the high luminosity $\gamma$-ray state. In the first model (Takata et al.~2014), the accretion disk does not penetrate below the light cylinder radius of the pulsar magnetosphere due to its evaporation by $\gamma$-ray radiation produced in the outer gap (Takata et al.~2010) or as a result of the disk interaction with the pulsar wind (Burderi et al.~2001). 
In the second model (Papitto et al.~2014b) proposes that the pulsar mechanism is quenched by the accretion disk which penetrates below the light cylinder radius. The $\gamma$-rays in the disk state are produced in the synchrotron self-Compton process by electrons accelerated on the border between the rotating magnetosphere and 
the inner accretion disk in a similar way as recently postulated for the accretion powered 
LMXBs (e.g. Bednarek~2009). Here we consider another scenario in which the accretion disk slowly penetrates the inner pulsar magnetosphere up to the co-rotation radius, allowing the matter to fall onto the NS surface. However, the acceleration gaps in the pulsar magnetosphere, which developed at large distance from the disk (e.g. expected in terms of the slot gap model), are not switched off. Due to the geometry of the magnetic field outside the light cylinder radius, the matter from the disk cannot penetrate into gaps at large distance from the disk surface. 
We propose that secondary leptons, produced in the region of the slot gap, comptonize disk radiation 
to GeV energies. We consider in detail the radiation processes which can be responsible for the 
high states of $\gamma$-ray emission in the best studied up to now MSP binary system PSR J1227-4853.

\section{A model for the high energy radiation in the disk state}  

In our modelling we concentrate on the redback binary system  containing millisecond pulsar 
PSR J1227-4853 for which the spectral information on the $\gamma$-ray emission in different states 
is precise enough for more detailed interpretation. The post-transition $\gamma$-ray emission from 
the redback binary system containing PSR J1227-4853 shows clear modulation
with the period of the pulsar (Johnson et al.~2015). Therefore, it is expected to be produced in terms of rotation powered pulsar mechanism. According to the outer gap and slot gap models this emission mostly comes from the region close to the light cylinder radius and originates in the curvature process of primary leptons accelerated in the gap to the Lorentz factors of the order of a few $10^6$.  

\begin{figure}
\vskip 5.6truecm
\includegraphics{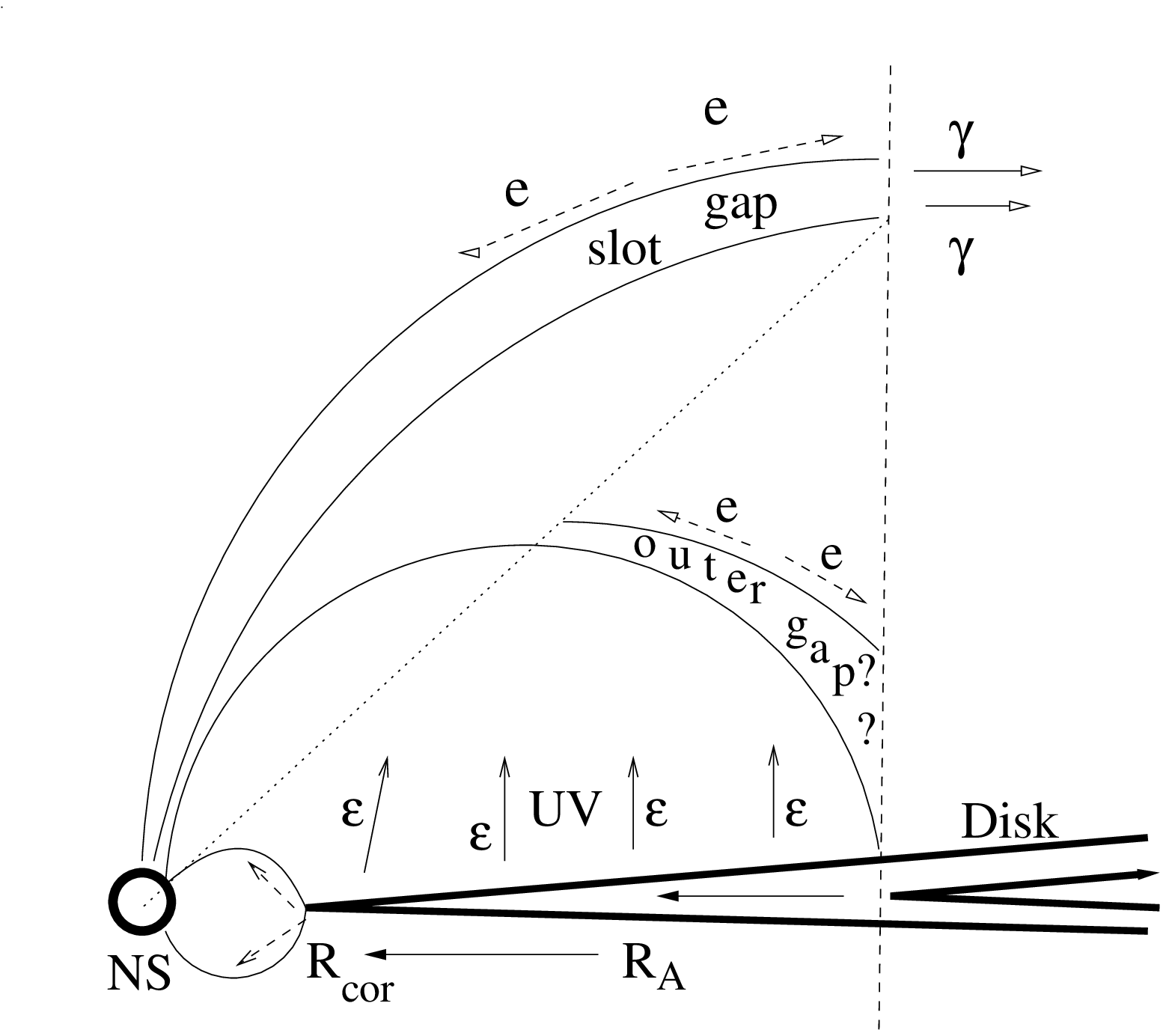}
\caption{Schematic representation (not to scale) of the vicinity of the millisecond pulsar within the binary system in which the accretion disk slowly builds up and gradually penetrates the inner millisecond pulsar magnetosphere. Due to the increased density of the matter at the inner radius of the disk, the Alfven radius $R_{\rm A}$  approaches the co-rotation radius $R_{\rm cor}$ in the pulsar magnetosphere. When these radii becomes comparable, the matter starts to fall onto the neutron star (NS) following local magnetic field lines. We expect that for pulsars not far from the alignment, the matter from the disk is not able to
penetrate into the parts of the pulsar magnetosphere at large distances from the rotational plane due to the toroidal geometry of the magnetic field above the light cylinder and the lack of plasma there. Therefore, the acceleration process of electrons and their radiation processes, typical for the slot gap or extended polar gap, should not be directly influenced by the presence of the accretion disk in the inner pulsar magnetosphere. It is not clear whether this is also the case of the outer gap model in which the acceleration region is much closer to the disk surface. 
In such scenario the lower level $\gamma$-ray emission in the rotation powered state comes from the primary electrons accelerated in the slot gap (curvature radiation) but the emission in the presence of the  accretion disk is the sum of this lower level emission from the gap and the $\gamma$-ray emission produced in the comptonization process of thermal disk radiation by secondary leptons close 
to the slot gap.   
}
\label{fig1}
\end{figure}

On the other hand, the $\gamma$-ray emission in the pre-transition state, in which the accretion disk is present, is proposed to be produced in another mechanism (see discussion in the Introduction). Here we consider another scenario for the pre-transition emission state in the binary PSR J1227-4853. In proposed scenario, the accretion disk can penetrate below the light cylinder radius but it does not switch off completely the pulsar mechanism (for schematic view of the scenario see Fig.~1). This can happen since in the case of nearly aligned rotator the plasma in the accretion disk is not directly connected by the magnetic field lines to the regions in the inner pulsar magnetosphere which are at large distances from the equatorial plane. In fact, the accretion disk might have a corona or a disk wind. It is expected  that the density of matter outside the accretion disk in such a wind or an extended corona is much rare than the density of matter within the disk. Therefore, its energy density is likely to be also lower than the energy density of the magnetic field in the pulsar wind zone. Note that the toroidal magnetic field above the light cylinder radius weakly depends on the distance from the NS. Such magnetic field might additionally stabilize the accretion disk against producing extended corona or the wind. The slot gap appears in the pulsar magnetosphere as a result of the space charge limited outflow scenario (Arons \& Scharleman~1979, Harding \& Muslimov~1998). In this model the matter from the NS surface is not expected to saturate the electric field of the gap since plasma with Goldreich \& Julian (1969, GJ) density at the NS surface drops below GJ density at some distance from the surface due to the dipole geometry of the magnetic field lines. Thus, the conditions for the appearance of the accelerating gap are satisfied.

Along the magnetic field lines above the null surface (see dotted line in Fig.~1), electrons can be accelerated in the slot gap scenario. As expected in classical pulsar model, we assume that the post-transition $\gamma$-ray emission is due to curvature radiation of the primary electrons accelerated in the electric field of the slot gap model (e.g. Arons~1983). A part of these primary curvature $\gamma$-rays is absorbed close to the gap and creates the
secondary population of $e^\pm$ plasma. In fact, the number of secondary leptons can be even 
significantly larger with the presence of the thermal X-ray radiation from the nearby accretion disk 
than in the case of clean pulsar magnetosphere. Additional disk radiation field can 
significantly increase the population of secondary leptons due to absorption of primary gamma-rays in the X-ray radiation as postulated by e.g. Romani~(1996).
These secondary leptons comptonize thermal radiation from the nearby accretion disk penetrating deep into the inner pulsar magnetosphere.
In such scenario, the pre-transition $\gamma$-ray emission state is the sum of the $\gamma$-ray emission produced in the inverse Compton process of secondary leptons and the lower level $\gamma$-ray emission produced in the pulsar mechanism (observed also in the post-transition state). 
We suppose that the production of $\gamma$-rays (and comptonization of the disk radiation by leptons) in terms of the the outer gap model (Cheng et al.~1986) may be impossible in such scenario. Note that the accretion disk, penetrating the inner pulsar magnetosphere, is quite close to the outer gap at the light cylinder radius. Thus, a part of the matter from the disk could easily penetrate into the outer gap switching off its electric field (see Fig.~1).

The transition states are expected to be initialized by the enhanced accretion rate from the stellar companion of the pulsar due to the Roche lobe overflow. The rotational axis of the pulsar should be close to alignment to the rotational axis of the binary system due to the past accretion history of the accreting matter. Therefore, the matter is expected to penetrate the outer pulsar magnetosphere along the rotational plane of the binary system. We assume that the pulsar is not far from alignment with small angle between the magnetic and rotational axis. In the transition state the pressure of the matter, accreting in the rotational plane, is not balanced by the pressure of the pulsar wind. Therefore, accretion disk starts to build up. With accumulation of matter in the accretion disk, its inner radius shrinks reaching at first the light cylinder radius and after that it can penetrate into the inner pulsar magnetosphere. The location of the disk inner radius in the pulsar magnetosphere (below the light cylinder) can be estimated by balancing the kinetic energy density of the plasma in the disk with the magnetic field energy density,
$B^2(R)/8\pi = \rho v_{\rm K}^2/2$,
where $B(R) = B_{\rm NS} (R_{\rm NS}/R)^3$ is the magnetic field strength at the distance $R$ from the centre of the NS, $v_{\rm K} = (GM_{\rm NS}/R)^{1/2}$ is the Keplerian velocity of the matter in the accretion disk, $R_{\rm NS}$ and $B_{\rm NS}$ are the radius of NS and its surface magnetic field strength, $\rho$ is the density of matter at the inner disk radius $R_{\rm in}$ (in grams). The above condition allows us to estimate the location of the so called Alfven radius, $R_{\rm A} = 8.4\times 10^4 (B_8^2/\rho)^{1/5}$ cm.
In fact, the inner disk radius is expected to be located at the magneto-spheric radius, $R_{\rm m}$, which is not precisely known fraction of the Alfven radius, $R_{\rm m} = \chi R_{\rm A}$. $\chi$ is expected to lay in the range $\chi\sim$0.1-1 (see e.g. Lamb, Pethick \& Pines~1973, and more recent discussion of this value in Bozzo et al.~(2009).
Even for relatively low density of accreting matter, the disk can penetrate the inner pulsar magnetosphere below the light cylinder radius. Note that the Alfven radius slowly shrinks with the increasing density of the matter in the inner disk. In this stage the accretion disk resembles 
the models of the dead disks (Sunyaev \& Shakura 1977) or trapped disks (Spruit \& Taam~1993, D'Angelo \& Spruit~2010, 2012) around compact objects in which substantial amount of matter can be accumulated but the thermal structure of the disk is determined by the continuously supplied fresh matter through the 1st Lagrangian point of the binary system. The process of building up of the disk continue up to the moment when the magneto-spheric radius reaches so called co-rotation radius defined as the distance from NS at which the Keplerian velocity of matter becomes comparable with the rotational velocity of the magnetic field lines,
$v_{\rm K} = v_{\rm rot}$, where $v_{\rm rot} = 2\pi R/P_{\rm NS}$, 
$P_{\rm NS} = 10^{-3}P_{\rm ms}$ s and $M_{\rm NS}$ are the pulsar rotational period and its mass assumed to be $1.4$ M$_\odot$. From this moment, the matter from the disk can fall onto the surface of NS. The co-rotation radius is located at, $R_{\rm cor} =  (GM_{\rm NS})^{1/3}(P_{\rm NS}/2\pi)^{2/3}\approx 1.7\times 10^6P_{\rm ms}^{2/3}$ cm.
For $R_{\rm m}\approx R_{\rm cor}$, the matter starts to fall onto the NS following local magnetic field lines. This distance defines also in our scenario the inner radius of the accretion disk. If the magneto-spheric radius
is always lower than the co-rotation radius then the disk extends up to the surface of NS. 

In the quasi-steady state, the matter flows through the accretion disk at the rate defined by the accretion rate of the matter passing through the 1st Lagrangian point.
We assume that this accretion energy is irradiated from the disk surface as expected in the Shakura \& Sunyaev~(1973) model.  Then, the disk luminosity is $L_{\rm D} = GM_{\rm NS}\dot{M}/2R_{\rm in} = 4\pi R_{\rm in}^2\sigma_{\rm SB}T_{\rm in}^4$, where $\dot{M}$ is the accretion rate. Based on this model, we estimate the characteristic temperature of the disk at the inner radius on,
$T_{\rm in}\approx  1.9\times 10^6L_{34}^{1/4}~~~{\rm  K}$, if the disk extends up to the NS surface (then $R_{\rm m} < R_{\rm cor}$) and on, $T_{\rm in}\approx 1.5\times 10^6L_{34}^{1/4}/P_{\rm ms}^{1/3}$ K, if the disk extends only to the co-rotation radius (then $R_{\rm m}\approx R_{\rm cor}$).
The disk luminosity, $L_{\rm D} = 10^{34}L_{34}$ erg s$^{-1}$, corresponds to the accretion rate onto the NS surface equal to $\dot{M}\approx 1.07\times 10^{13}$ g s$^{-1}$. 

The radiation field in the slot gap has been calculated assuming the Sunyaev \& Shakura disk model in which 
the gravitational energy extracted by the plasma is locally irradiated from the disk surface. In this model
the temperature profile on the disk surface is given by approximate formula $T(R)\approx T_{\rm in} (R_{\rm in}/R)^{3/4}$. We calculate the density of photons in the slot gap by integrating over the disk temperature profile following prescription given in e.g. Frank et al.~(1985, see Eq.~5.43). 
The maximum power is radiated from the surface with temperature $T_{\rm in}$ at photon energies 
$\varepsilon\sim 3k_{\rm B}T_{\rm in}\approx 390(L_{34}/P_{\rm ms}^3)^{1/4}$ eV. For the disk luminosity 
$3\times 10^{33}$ erg s$^{-1}$ and the pulsar period 1.7 ms is $\sim 200$ eV, falling into the UV energy range.
Such photons are comptonized on the border between the Thomson (T) and the Klein-Nishina (KN) regimes by electrons with energies $E_{\rm T/KN} = m_{\rm e}^2/\varepsilon\sim 0.7(P_{\rm ms}^3/L_{34})^{1/4}$ GeV. Therefore, 
in the calculations of the IC spectra, the KN effects has to be taken into account.
The mean free path for electrons (in the Thomson regime) for scattering thermal radiation from the disk with characteristic temperature $T_{\rm in}$ is,
$\lambda_{\rm IC/T} = (\sigma_{\rm T}n_{\rm ph})^{-1}\approx 2.7\times 10^5P_{\rm ms}^{5/3}/L_{34}^{3/4}$ cm,
where $n_{\rm ph}\approx 380(T_{\rm in}/2.7 K)^3(R_{\rm in}/R_{\rm LC})^2$ ph cm$^{-3}$ is the density of thermal photons from the disk in the location of the slot gap close to the light cylinder radius, and $\sigma_{\rm T}$ is the Thomson cross section. This distance scale is smaller than the distance scale for the propagation of secondary leptons in the inner pulsar magnetosphere given by the light cylinder radius for the disk luminosity above, $L_{\rm min}\approx 2.2\times 10^{32}P_{\rm ms}^{8/9}$ erg s$^{-1}$.
We also calculate the optical depths for absorption of curvature $\gamma$-rays, produced by primary electrons, and the IC $\gamma$-ray spectra, produced by secondary leptons, in the radiation field of the accretion disk. The absorption of $\gamma$-rays strongly depend on the accretion rate (and so the disk luminosity). Since the $\gamma$-ray production region in the slot gap is assumed to be homogeneous, with the characteristic dimension equal to $R_{\rm LC}$, the absorption effects are approximately included by multiplying produced $\gamma$-ray spectra by the factor equal to $(1 - exp(-\tau))/\tau$. We show the example calculations of the $\gamma$-ray spectra produced in terms of such model including the effects of their absorption (Fig.~2). The parameters of the MSP binary system PSR J1227-4863 are assumed. Produced IC spectra are calculated for different luminosities of the accretion disk. 
It is clear that absorption of $\gamma$-rays starts to be important for the disk luminosities above $\sim 10^{33}$ erg
s$^{-1}$. In the next section we apply such scenario for the $\gamma$-ray production in the pre-transition state of the redback type MSP binary system PSR J1227-4853 for which the spectra in different states have been well measured (Xing \& Wang~2014, Johnson et al.~2015). 

\begin{figure}
\vskip 4.2truecm
\includegraphics{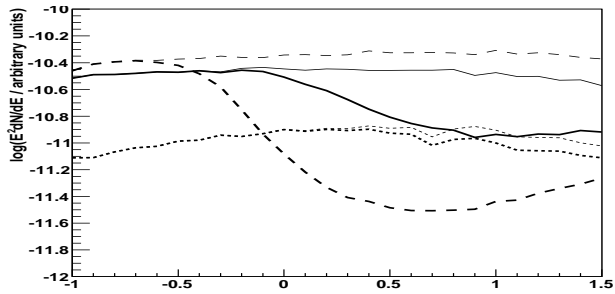}
\caption{Spectral Energy Distribution (SED, differential spectra multiplied by energy squared) of 
$\gamma$-rays produced in the IC scattering of the accretion disk radiation by secondary leptons with the power law spectrum and spectral index equal to -2 between 100 MeV and 100 GeV (thin curves) and the spectra after absorption of $\gamma$-rays in the disk radiation (thick curves). Specific spectra are calculated for luminosities of the accretion disk $L_{\rm D} = 10^{33}$ erg s$^{-1}$ (dotted), $10^{34}$ erg s$^{-1}$ (solid), and $10^{35}$ erg s$^{-1}$ (dashed). The disk extends up to the co-rotation radius and the period of the pulsar is equal to 1.7 ms.}
\label{fig2}
\end{figure}
\section{Gamma-ray emission states in binary system PSR J1227-4853}

The baseline of the considered scenario is the assumption that enhanced $\gamma$-ray emission in the pre-transition state in PSR J1227-4853 is produced by the secondary $e^\pm$ leptons which up-scatter disk thermal radiation to GeV energies. In fact, $e^\pm$ pairs should be present in the region of the slot gap of the pulsar since in the post-transition state a clear radio pulsations are observed. Such coherent radio emission has to originate in dense bunches of 
e$^\pm$ plasma. The $\gamma$-ray spectrum in the post-transition state is expected to originate in the curvature process of primary electrons accelerated in the electric field of the slot gap. We assume that a large amount of secondary leptons is produced with close to the power law spectrum
in the energy range between 100 MeV and 100 GeV. In fact, such type of secondary leptons spectra are expected to be produced in the pulsar magneto-spheres in the case of the existence of the extended polar cap regions (see e.g. Hibschman \& Arons~2001). 
In the pre-transition state, the accretion disk penetrates deep into the inner pulsar magnetosphere providing additional soft photon target to be up-scattered by secondary leptons. We simulate the propagation of leptons on the distance scale of the light cylinder radius in the thermal radiation field from the inner accretion disk. The synchrotron energy losses of leptons are taken into account assuming  their average pitch angle of of the order of 0.1 rad. For the magnetic field strength of the NS at the light cylinder radius, the synchrotron energy losses of leptons are estimated to be clearly lower than their energy losses on the IC process in the accretion disk radiation field.

The $\gamma$-ray spectrum in the pre-transition state is reasonably well described by the sum of the spectrum from the post-transition state (curvature radiation of primary electrons) and the spectrum produced by secondary leptons in the inverse Compton scattering of disk radiation. This fitting to the pre-transition $\gamma$-ray spectrum has been obtained for the simple assumption on the spectrum of secondary leptons which has been assumed to be of the power law type with the spectral index equal to -2.6 between 100 MeV and 100 GeV and the disk radiation field which luminosity is equal to $3\times 10^{33}$ erg s$^{-1}$ (see~Fig.~3). In order to obtain correct $\gamma$-ray luminosity in the pre-transition state, the power in secondary leptons should be equal to $\sim$0.33 of the product of the the pulsar spin down luminosity (equal to $9\times 10^{34}$ erg s$^{-1}$ in the case of PSR J1227-4853) and the collimation factor of the $\gamma$-ray emission produced by leptons in the slot gap.

Since the secondary leptons have some degree of beaming it is possible that the $\gamma$-ray emission during pre-transition state can also show some modulation with the period of the pulsar. However the presence of the accretion disk can also influence the pulsar characteristic spin down rate which might complicate direct detection of such modulation. It is also not clear whether the radio pulsar should be still visible in the pre-transition state since the presence of the additional radiation from the accretion disk in the inner magnetosphere might influence of formation of bunches responsible for the coherent radio emission or the accretion disk can simply screen the region responsible for the radio emission.

\begin{figure}
\vskip 4.3truecm
\includegraphics{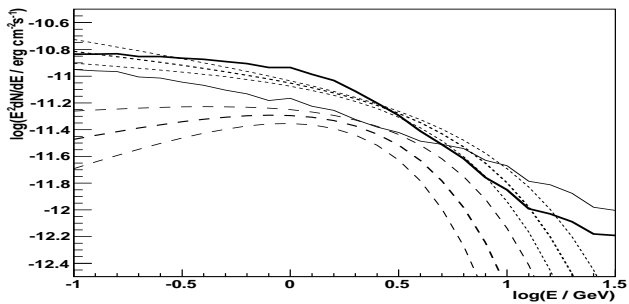}
\caption{Spectral Energy Distribution (SED) of the Redback type binary system containing MSP PSR J1227-4853. The approximation of the pre-transition and post-transition spectra from PSR J1227-4853 (dotted and dashed curves, see Johnson et al.~2015). The IC spectrum produced by secondary $e^\pm$ pairs which comptonize thermal radiation from the accretion disk, before (thin solid curve) and after absorption in the disk radiation (thick solid). It is assumed that the secondary e$^\pm$ spectrum is of a simple power law type with the spectral index -2.6 between 0.1-100 GeV. The accretion disk luminosity is assumed to be equal to $L_{\rm D} = 3\times 10^{33}$ erg s$^{-1}$.}
\label{fig3}
\end{figure}
\section{Conclusion}

Observations of the pulsed X-ray emission from millisecond pulsar binary systems, in the state with the presence of an accretion disk, suggest that the disk has to penetrate deep into the inner pulsar magnetosphere and that at least some of the disk matter has to fall onto the neutron star surface. Based on this evidence we propose that high $\gamma$-ray emission state from MSP binary systems, observed in the presence of the accretion disk, is produced
in the comptonization process of the disk radiation by secondary leptons created relatively far from the disk in the region of the slot gap, i.e. above the null surface in the MSP magnetosphere.  In contrast to the outer gap scenario, $e^\pm$ pairs in the slot gap, are expected to be produced copiously since that acceleration region is far away from the disk. It is not expected to be quenched by the dense plasma from the disk. In this scenario, $\gamma$-rays in the rotation powered state originate in the curvature process of primary electrons.
In the disk state, primary electrons produce curvature $\gamma$-rays, which can be additionally absorbed in the disk radiation ejecting secondary leptons in the cascade process occurring above the null surface. These secondary leptons are expected to have a few orders of magnitude lower energies than primary electrons. Therefore, they are able to scatter efficiently UV to soft X-ray radiation produced by the accretion disk in contrast to primary electrons.

We simulate the propagation and radiation process of secondary leptons, in the region close to the light cylinder radius and above the null surface. The accretion disks with luminosities in the range expected for the accreting millisecond pulsars are considered. In our calculations we take into account IC scattering of disk radiation, synchrotron losses of secondary leptons and also absorption of $\gamma$-rays in the disk radiation. As an example, the results of calculations of the IC $\gamma$-ray spectra are compared with the two $\gamma$-ray emission states from the transiting MSP Redback type binary system J1227-4853 for which the spectral information in the $\gamma$-ray energy range has been precisely measured (Johnson et al.~2015). Reasonable description has been obtained in terms of such model for likely parameters of this system. 

In this model the $\gamma$-ray production process in the accretion disk state is determined 
by the active pulsar mechanism. Therefore this scenario does not predict significant $\gamma$-ray emission from the powerful LMXBs containing slowly rotating, but strongly magnetised, neutron stars in contrast to the scenario for the acceleration and radiation processes in the inner region of the accretion disk recently considered for such LMXBs (Bednarek~2009) and also for the millisecond pulsars (Papitto et al.~2014b).

Another transitional millisecond pulsar, PSR J1023+0038, shows the flux change in the $\gamma$-rays by a factor of $\sim$5 in respect to the change by a factor of $\sim$2 in the case of discussed here source. We think that such large flux changes between different states could be also understood in terms of the proposed scenario. In principle the power in secondary leptons, responsible for additional component in the disk state, can be larger than the power in observed $\gamma$-rays. The primary $\gamma$-rays can be efficiently absorbed in the gap region and only a small part can survive this process and escape to the observer. These leptons meet strong radiation field from the accretion disk allowing them efficient conversion of their energy to the IC $\gamma$-ray component observed in the disk state of
transitional pulsars. Moreover the emission geometry can also play an important role. The rotating disk plasma is expected to produce additional component of the magnetic field which might slightly influence the dipole component of 
NS. As a result, the structure of magnetosphere at the light cylinder can change allowing enhanced $\gamma$-ray emission at certain directions.

\section*{Acknowledgments}
I would like to thank the Referee for useful comments which allowed to improve the paper.
This work is supported by the grant through the Polish Narodowe Centrum Nauki No. 2011/01/B/ST9/00411.


\label{lastpage}
\end{document}